\documentclass[12pt,preprint]{aastex}
\begin{document}
\title{Collision of two identical hypersonic stellar winds in
binary systems}
\author{Nikolay N. Pilyugin~\altaffilmark{1} and Vladimir V.~Usov}
\affil{Center for Astrophysics, Weizmann Institute,
Rehovot 76100, Israel} \altaffiltext{1}{Institute of Mechanics,
Moscow State University,
Mitchurinsky prosp. 1, Moscow, 111531, Russia}
\begin{abstract}
We investigate the hydrodynamics of two identical hypersonic stellar
winds in a binary system. The interaction of these winds manifests
itself in the form of two shocks and a contact surface between them.
We neglect the binary rotation and assume that the gas flow ahead
of the shocks is spherically symmetrical.
In this case the contact surface that separates
the gas emanated from the different stars coincides with the
midplane of the binary components. In the shock the gas is heated and flows
away nearly along the contact surface. We find the shock shape and
the hot gas parameters in the shock layer between the shock and
the contact surface.
\end{abstract}
\section{Introduction}
Stars of various types drive stellar winds with widely differing
characteristics. In binary systems that consist of such stars the
winds flowing out of the stars may strike each other. Collision of
stellar winds may result in an observational appearance if the
winds are rather powerful. Among all stars, massive OB and
Wolf-Rayet (WR) stars have the strongest winds. The mass-loss rate
$\dot M$ is as high as $\sim 10^{-6}-10^{-5}\,M_\odot {\rm
yr}^{-1}$ for O and B stars or even higher ($\sim
10^{-5}-10^{-4}\,M_\odot {\rm yr}^{-1}$) for WR stars. The
terminal velocity of the matter outflow is $V^\infty\sim
(1-5)\times 10^3$ km~s$^{-1}$ for OB and WR stars. The kinetic
energy carried away by the winds is $\sim 10^{35}-10^{38}$
ergs~s$^{-1}$. In the region where the two winds collide some part
of this energy may be transformed to other forms of energy (for
example, the thermal energy of hot gas or the energy of
ultra-relativistic electrons), and then, radiated at different
frequencies. At present a variety of phenomena related to the
collision of the stellar winds are observed for WR+OB and OB+OB
binaries (for a review, see Williams 1999; Cherepashchuk 2000;
Corcoran et al. 2005). In particular, it is found that the X-ray
luminosities of massive binaries are significantly higher (by a
factor of $\sim 10^2$ for some systems) than the total X-ray
luminosities of their single counterparts in a statistical sense
(Pollock 1987; Chlebowski \& Garmany 1991). The analysis of the
energy spectra and the time variability of the X-ray emission
indicates that the X-ray excess is emitted from the region where
the two winds collide (Moffat et al. 1982; Chlebowski 1989;
Williams et al. 1990; Corcoran et al. 1996; Corcoran 2003). The
observational data on the X-ray emission from massive binaries are
well consistent with the model where strong shocks form because of
the wind-wind collision (Prilutskii and Usov 1975, 1976;
Cherepashchuk 1976; Usov 1990, 1992, 1995; Stevens, Blondin, \&
Pollock 1992; Antokhin, Owocki, \& Brown 2004). In this model the
outflowing gas is heated behind of the shocks to temperatures of
$\sim 10^6-10^7$~K and radiate X-ray emission that is observed as
the X-ray excess. Besides, the shock waves in massive binaries are
able to accelerate electrons to ultrarelativistic energies, and
these electrons can generate non-thermal radio emission via the
synchrotron mechanism in the vicinity of the shocks (Williams et
al. 1990; Eichler \& Usov 1993; Williams, van der Hucht, \&
Spoelstra 1994; White \& Becker 1995; Dougherty et al. 1996;
Jardine, Allen, \& Pollock 1996; Chapman et al. 1999; Setia
Gunawan et al. 2000; Monnier et al. 2002; Pittard et al. 2006).
For several nearby WR+OB binaries (WR140, WR146 and WR147), the
radio sources are spatially resolved, and the radio emission model
based on the wind-wind collision is confirmed (Moran et. al. 1989;
Churchwell et al. 1992; Niemela et al. 1998; Dougherty, Williams,
\& Pollacco 2000) Dust formation may also occur in the wind
collision region (Usov 1991) and is observed for several massive
binaries as an IR excess (Williams \& van der Hucht 1992;
Marchenko, Moffat, \& Grosdidier 1999; Monnier, Tuthill, \& Danchi
2002; Williams et al. 2003).

The hydrodynamics of colliding winds has been studied both
analytically (Galeev, Pilyugin, \& Usov 1989; Bairamov, Pilyugin,
\& Usov 1990; Usov 1992, 1995) and numerically (Luo, McCray, \&
Mac Low 1990; Myasnikov \& Zhekov 1991; Stevens, Blondin, \&
Pollock 1992; Pittard \& Stevens 1997; Zhekov \& Skinner 2000;
Henley, Stevens, \& Pittard 2003). In the analytical studies it
was assumed that the wind of star 1 (the primary) is much stronger
than that of star 2 (the secondary). In this case the wind-wind
collision zone is near the secondary on the side facing the
primary, and its characteristic size is much smaller than the
distance between the components of the binary. The undisturbed gas
stream from star 1 ahead of the shock was considered as
plane-parallel. The wind from star 2 was assumed to be spherically
symmetrical. It is a reasonable approximation, for instance, to
the WR+OB binaries where the wind momentum of the OB star is $\sim
10-10^3$ times smaller than that of the WR star, i.e., $\eta =\dot
M_{\rm OB}V_{\rm OB}^\infty/\dot M_{\rm WR}V_{\rm WR} ^\infty\sim
10^{-3}-10^{-1}\ll 1$. However, this approximation is too rough
for binaries with nearly the same stellar winds ($\eta\sim 1$). In
this paper we consider the wind-wind collision in the case when
the colliding winds are identical in strength ($\eta =1$).

The paper is organized as follows. In \S~2 we discuss the properties
of the undisturbed gas that flows away from the star vicinity.
In \S~3 we formulate the problem of collision of two identical
hypersonic steady winds in a binary system and outline both the
hydrodynamic equations, which describe the motion
of the hot gas behind the shock, and boundary conditions on them.
In \S~4 we solve the set of hydrodynamic equations and boundary
conditions and find the structure of the shock and the parameters
of the hot gas in the shock layer. Finally, in \S~5 we discuss our
results and some potential astrophysical applications.

\section{Formulation of the problem}

We consider the hydrodynamics of two identical hypersonic steady
winds in a binary system. We assume that the gas flow from the
binary components is spherically symmetrical, nonviscous and
nonheat conductive. The energy losses of the gas by radiation are
neglected, i.e., the gas flow is adiabatic. We also neglect the
rotation of the binary and its components.

The interaction of two colliding stellar winds will manifest
itself in the form of two shocks separated by a contact surface
(see Fig.~1). The winds from the binary components flow radially
out to the shocks. In the shock the gas is heated to the postshock
temperature of $\sim 10^7V_8$~K, where $V_8$ is the preshock wind
velocity perpendicular to the shock in units of $10^8$
cm~s$^{-1}$. Behind the shock the hot gas flows away from the
binary system nearly along the contact surface (e.g., Usov 1992
and below). The contact surface separates the hot gas that has
been emanated from the different components. For an arbitrary
value of $\eta$ the contact surface has a rather complex shape
(e.g., Canto et al. 2006). In our case of two equal winds
($\eta=1$), there is a reflection symmetry about the midplane of
the two stars, and the contact surface coincides with the midplane
(see Fig.~1).

\subsection{Basic equations}

We start from the set of equations, which describes the gas flow
in our approximation,
\begin{equation}
{\rm div}\,(\rho{\bf V})=0\,,
\label{basiceq1}
\end{equation}
\begin{equation}
\rho({\bf V}{\bf \nabla}){\bf V} = - {\bf \nabla} p\,,
\label{basiceq2}
\end{equation}
\begin{equation}
\rho{\bf V}{\bf \nabla} (H + |{\bf V}|^2/2) =0\,,
\label{basiceq3}
\end{equation}
where ${\bf V}$ is the velocity of the gas, $\rho$ is its density,
$p$ is the total pressure,
\begin{equation}
H={\gamma\over {\gamma-1}}{p\over \rho}
\label{entH}
\end{equation}
is the specific enthalpy, $T$ is the temperature, and $\gamma$ is the
ratio of heat capacities at constant pressure
and at constant volume
(e.g., Chernyi 1961). Equation (\ref{basiceq1}) gives
conservation of mass, while equations (\ref{basiceq2}) and
(\ref{basiceq3}) conserve the momentum and total energy, respectively.

In the shock layers between the shocks and the contact surface the
hot gas may be considered as a rarefied totally ionized plasma.
For such a plasma, we have
\begin{equation}
p =(N_e+N_i)kT= {\rho kT\over m_p\mu}\,,
\label{ptotal}
\end{equation}
\begin{equation}
N_i={\rho\over  m_pA}\,,\,\,\,\,\,
N_e= N_iZ\,,\,\,\,\,\,
{\rm and}\,\,\,\,\,
\gamma={5\over 3}\,,
\label{ngamma}
\end{equation}
where $N_e$ is the density of electrons, $N_i$ is the density of
ions, $k$ is the Boltzmann
constant, $A$ is the atomic weight of ion,
$Z$ is its electrical charge, $\mu=A/(1+Z)$ is the mean molecular
weight, and $m_p$ is the proton mass.

\subsection{Boundary conditions}
The gas parameters ahead of the shock (index 1) and behind the
shock (index 2) are related via the Rankine-Hugoniot relations
\begin{eqnarray}
\rho_1V^{(n)}_1=\rho_2V^{(n)}_2, \,\,\,
p_1+\rho_1[V^{(n)}_1]^2=p_2+\rho_2 [V_2^{(n)}]^2,  \nonumber\\
V_1^{(\tau)}=V_2^{(\tau)},\,\,\,\, H_1+{[V_1^{(n)}]^2\over
2}=H_2+{[V_2^{(n)}]^2\over 2}\,.\,\,\,\, \label{RH}
\end{eqnarray}
Indices $n$ and $\tau$ denote the normal and tangential components of the
gas velocity ${\rm V}$.  The condition $V^{(n)}$=0 is met on the contact
surface.  This condition and the Rankine-Hugoniot relations (\ref{RH})
are the
full set of boundary conditions for the set of equations
(\ref{basiceq1})-(\ref{ptotal}) needed
to find the parameters of
hot gas in the shock layers.

\section{Stellar winds from massive stars}
The luminosity  of a massive star is very high, and the radiation
pressure is responsible for the outflow of gas and its subsequent
acceleration. The gas velocity $V(r)$ varies from almost zero at
the surface of the star ($r=R)$ to some terminal value at the
distance $r_*\simeq (3-5)R$ from the stellar center (e.g., Barlow
1982).

At $r\gtrsim r_*$ the gas acceleration is more or less finished,
and the gravity of the star may be neglected. In this case,
from equations (\ref{basiceq1})-(\ref{entH})
the spherically symmetric supersonic (${\mathfrak M}>1$)
flow of gas may be described by
\begin{equation}
\left({r_*\over r}\right)^2=\left({{\mathfrak M}\over {\mathfrak M}_*}
\right)\left[{(\gamma-1){\mathfrak M}^2_*+2\over (\gamma-1){\mathfrak M}^2
+2}\right]^{\gamma+1\over 2(\gamma-1)},
\label{eqwind1}
\end{equation}
\begin{equation}
{V\over V_*}=\left({{\mathfrak M}\over {\mathfrak M}_*}
\right)\left[{(\gamma-1){\mathfrak M}^2_*+2\over (\gamma-1){\mathfrak M}^2
+2}\right]^{1/2},
\label{eqwind2}
\end{equation}
\begin{equation}
{\rho\over \rho_*}=\left[{(\gamma-1){\mathfrak M}^2_*+2\over
(\gamma-1){\mathfrak M}^2
+2}\right]^{1\over \gamma-1},
\label{eqwind3}
\end{equation}
\begin{equation}
{p\over \rho_*V_*^2}={1\over \gamma\, {\mathfrak M}^2_*}
\left[{(\gamma-1){\mathfrak M}^2_*+2\over (\gamma-1){\mathfrak M}^2
+2}\right]^{\gamma\over \gamma-1},
\label{eqwind4}
\end{equation}
\begin{equation}
{2H\over V_*^2}= {2\over (\gamma-1)\, {\mathfrak M}_*^2}
\left[{(\gamma-1){\mathfrak M}^2_*+2\over (\gamma-1){\mathfrak M}^2
+2}\right]\,,
\label{eqwind5}
\end{equation}
where ${\mathfrak M}=V/V_s$ is the Mach number and $V_s$ is the sound
speed in the outflowing gas. The gas parameters marked by the sign
$_*$ are
taken at $r=r_*$. For winds from massive stars, we have $V\sim 10^3$
km~s$^{-1}$, $V_s\sim 10$ km~s$^{-1}$, and ${\mathfrak M}\sim 10^2\gg 1$.

For $(\gamma-1){\mathfrak M}^2_*\gg 1$  equations
(\ref{eqwind1})-(\ref{eqwind5}) are simplified and yield
\begin{equation}
{\mathfrak M}={\mathfrak M}_*\left({r\over r_*}\right)^{\gamma-1}
\gg 1\,,
\label{eqwind21}
\end{equation}
\begin{equation}
V=V_*=V^\infty,\,\,\,\,\,\,\,{\rho\over \rho_*}=\left({r_*\over r}
\right)^2,
\label{eqwind22}
\end{equation}
\begin{equation}
{p\over \rho_*V_*^2}={1\over \gamma\, {\mathfrak M}^2_*}
\left({r_*\over r}\right)^{2\gamma}\ll 1\,,
\label{eqwind23}
\end{equation}
\begin{equation}
{2H\over V_*^2}= {2\over (\gamma-1)\, {\mathfrak M}_*^2}
\left({r_*\over r}\right)^{2(\gamma-1)}\ll 1\,.
\label{eqwind24}
\end{equation}
From equations (\ref{eqwind23}) and (\ref{eqwind24}) we can see
that in a supersonic flow far from the star ($r\gtrsim r_*$)
the effects of
the thermal pressure and the thermal energy are of the order
of ${\mathfrak M}^{-2}$ that is $\sim 10^{-4}$ or less
for massive stars.
In our study, we neglect these effects, and the gas flow
ahead of the shocks is described by
\begin{equation}
{\bf V}= V^\infty{{\bf r}\over r}\,,\,\,\,\,\,\,\,
\rho=\rho_*\left({r_*\over r}\right)^2\,,\,\,\,\,\,
p=H=0\,.
\label{eqwfin}
\end{equation}

\section{Collision of two equal winds with terminal velocities}
In Section~2 it is noted that for two colliding stellar winds
the gas flow has reflection symmetry
about the contact surface that coincides with the midplane
of the binary components (see Fig.~1). Therefore, the problem of
collision of two equal winds is identical with the problem of collision
of one of the winds with the midplane. The latter problem is
considered in this paper. We assume that the binary is wide
($D>r_*$), and the terminal velocity is reached by the wind
ahead of the shock, where $D$ is a half of the binary separation.

\subsection{Shock layer}
In the shock layer the position of a point may be defined by the
two coordinates $x$ and $y$ that are the distances from the point
to the binary axis and the contact plane, respectively (see Fig.~2).
However, it is easier to solve the set of equations
(\ref{basiceq1})-(\ref{entH}) using
the independent variables $\psi$ and $x$, where $\psi$ is the stream
function defined by the equality
\begin{equation}
d\psi = \rho uxdy - \rho vxdx\,,
\label{psi}
\end{equation}
where $u$ and $v$ are the velocity components in the
$x$ and $y$ directions, respectively.

In the new independent variables $\psi$ and $x$ the set of
equations (\ref{basiceq1})-(\ref{entH}) can be written as
\begin{equation}
{\partial y\over\partial\psi} = {1\over\rho ux} , \; \; \; \; {\partial
y\over\partial x}= {v\over u} \ ,
\label{equfinal1}
\end{equation}
\begin{equation}
{\partial v\over \partial x} =- x{\partial
p\over\partial\psi} \ ,
\label{equfinal2}
\end{equation}
\begin{equation}
\rho\bigg (u{\partial u\over\partial x} + v{\partial v\over \partial
x}\bigg ) = - {\partial p\over\partial x} \ ,
\label{equfinal3}
\end{equation}
\begin{equation}
{\partial\over\partial x}\bigg(H+{u^2+v^2\over 2}\bigg) = 0\,.
\label{equfinal4}
\end{equation}

It is convenient to introduce dimensionless variables via
\begin{eqnarray}
\overline{x}={x\over D}\,,\,\,\,\,\,\,\,\overline{y}={y\over D}\,,
\,\,\,\,\,\,\,\overline{r}={r\over D}\,,\nonumber\\
\overline{u}={u\over V^\infty}\,,\,\,\,\,\,\,\,
\overline{v}={v\over V^\infty}\,,\nonumber\\
\overline{p}={p\over \rho_*(V^\infty)^2}\left({D\over r_*}
\right)^2,\,\,\,\,\,\,\,
\overline{H}={2H\over (V^\infty)^2}\,,\nonumber\\
\overline{\rho}={\rho\over \rho_*}\left({D\over r_*}\right)^2,
\,\,\,\,\,\,\,\,
\overline{\psi}={\psi\over \rho_*V^\infty r_*^2}\,.
\label{dimentionless2}
\end{eqnarray}
In these variables equations (\ref{equfinal1})-(\ref{equfinal4})
take the form:
\begin{equation}
{\partial \overline{y}\over\partial\overline{\psi}} =
{1\over\overline{\rho}\, \overline{u}\,\overline{x}} , \; \; \; \; {\partial
\overline{y}\over\partial \overline{x}}= {\overline{v}
\over \overline{u}} \ ,
\label{equfinaldl1}
\end{equation}
\begin{equation}
{\partial \overline{v}\over \partial \overline{x}}
=- \overline{x}{\partial
\overline{p}\over\partial\overline{\psi}} \ ,
\label{equfinaldl2}
\end{equation}
\begin{equation}
\overline{\rho}\bigg (\overline{u}{\partial \overline{u}
\over\partial \overline{x}} +
\overline{v}{\partial \overline{v}\over \partial
\overline{x}}\bigg )
= - {\partial \overline{p}\over\partial \overline{x}} \ ,
\label{equfinaldl3}
\end{equation}
\begin{equation}
{\partial\over\partial \overline{x}}
\bigg(\overline{H}+\overline{u}^2+\overline{v}^2\bigg) = 0\,.
\label{equfinaldl4}
\end{equation}

Using the boundary conditions (\ref{RH}) at the shock
(see Fig.~2) and equation (\ref{eqwfin}) for the gas parameters near
and ahead of the shock, we can get the gas parameters near and
behind the shock (index~$s$):
\begin{equation}
\overline{u}_s=\epsilon\,\cos\alpha\,\sin(\alpha+\varphi)
-\cos(\alpha+\varphi)\,\sin \alpha\,,
\label{behidshock1}
\end{equation}
\begin{equation}
\overline{v}_s=-[\epsilon\,\sin\alpha\,\sin(\alpha+\varphi)
+\cos(\alpha+\varphi)\,\cos\alpha]\,,
\label{behidshock2}
\end{equation}
\begin{equation}
\overline{p}_s={1\over \overline{r}_s^2}(1-\epsilon)\,
\sin^2(\alpha+\varphi)\,,
\label{behidshock3}
\end{equation}
\begin{equation}
\overline{H}_s=(1-\epsilon^2)\,\sin^2(\alpha+\varphi)\,,
\label{behidshock4}
\end{equation}
where $\epsilon$ is the ratio of the gas density ahead of the shock to that
behind it,
\begin{equation}
\epsilon ={\gamma -1\over \gamma+1}\,,
\label{behidshock5}
\end{equation}
$\alpha$ is the angle between the tangent to the shock
wave and the binary axis, $\varphi$ is the angle between the
radius-vector ${\bf r}$ from the center of the star and the binary axis,
and
\begin{equation}
\overline{r}_s=[\overline{x}^2+(1-\overline{y}_s)^2]^{1/2}
\label{rs}
\end{equation}
is the dimensionless distance from the stellar center to the shock
wave (see Fig.~2). For $\gamma={5\over 3}$, we have $\epsilon={1\over 4}$.

The equation of the shock shape
\begin{equation}
\overline{y}=\overline{y}_s(x)
\label{shockshape}
\end{equation}
is connected with the angle $\alpha$ by
\begin{equation}
\overline{y}_s'\equiv
{d\overline{y}_s\over d\overline{x}}= \tan^{-1}\alpha\,.
\label{shockshapealpha}
\end{equation}

From equations (\ref{eqwfin}) and (\ref{psi}) the stream function
at the shock wave may be written as
\begin{equation}
\overline{\psi}_s={\int_0^{\overline{x}}}
{\sin\,(\alpha+\varphi)\,\overline{x}d\overline{x}
\over \overline{r}^2_s}\,.
\label{psishock}
\end{equation}
In the dimensionless coordinates $\overline{x}$ and $\overline{\psi}$
the equation of shock shape
is $\overline{\psi}=\overline{\psi}_s(\overline{x})$.

Using equations (\ref{rs}) and (\ref{shockshapealpha}), equation
(\ref{psishock}) can be rewritten as
\begin{equation}
\overline{\psi}_s={\int_0^{\overline{x}}}
{(1+\overline{x}\,\overline{y}'_s-\overline{y}_s)
\,\overline{x}d\overline{x}
\over
[1+(\overline{y}'_s)^2]^{1/2}[\overline{x}^2+(1-\overline{y}_s)^2]
^{3/2}}\,.
\label{psishock2}
\end{equation}

Near the contact plane the velocity component perpendicular to
this plane is zero, i.e.,
\begin{equation}
\overline{v}=0\,\,\,\,\,\,\,{\rm at}\,\,\,\overline{y}=0\,.
\label{boundcondcont}
\end{equation}

To find the shock shape (\ref{shockshape}) and the parameters of
the hot gas in the shock layer we solve the set of equations
(\ref{entH}) and (\ref{equfinaldl1})-(\ref{equfinaldl4}) with the
boundary conditions (\ref{behidshock1})-(\ref{behidshock4}) and
(\ref{boundcondcont}) by the method of Chernyi (1961) in which
$\epsilon$ is considered as a small parameter. Henceforth, we omit
the bars over the dimensionless variables.

We seek a solution of equations
(\ref{equfinaldl1})-(\ref{equfinaldl4}) in
the form of a series in powers of $\epsilon$ in the following form:
\begin{equation}
y=\epsilon y_0 \,,\,\,\,\,\, u^2=u_0^2+\epsilon u_1^2
\,,\,\,\,\,\,
v=\epsilon v_0\,,
\label{series1}
\end{equation}
\begin{equation}
p=p_0+\epsilon  p_1\,,\,\,\,\,
\rho={\rho_0\over \epsilon}+\rho_1\,,\,\,\,\,
H=H_0+\epsilon H_1\,.
\label{series2}
\end{equation}
Substituting these series into equations (\ref{entH}) and
(\ref{equfinaldl1})-(\ref{equfinaldl4})
and equating the terms with the same power of
$\epsilon$, we obtain the following equations
for determining the terms of the series:
\begin{equation}
{\partial u_0\over \partial x}=0\,,\,\,\,\,\,\,
{\partial p_0\over \partial \psi}=0\,,\,\,\,\,\,\,
{\partial H_0\over \partial x}=0\,,
\label{equationsser1}
\end{equation}
\begin{equation}
v_0=u_0{\partial y_0\over \partial x}\,,\,\,\,\,\,\,
{\partial y_0\over \partial \psi}=
{1\over \rho_0x(u_0^2+\epsilon u_1^2)^{1/2}}\,,
\label{equationsser2}
\end{equation}
\begin{equation}
H_0={2\gamma\over (\gamma+1)}{p_0\over \rho_0}\,,\,\,\,\,\,\,
H_1=H_0\left({p_1\over p_0}-{\rho_1\over \rho_0}\right)\,,
\label{equationsser3}
\end{equation}
\begin{equation}
{\partial v_0\over \partial x}=-x{\partial p_1\over\partial \psi}\,,
\,\,\,\,\,\,
{\rho_0\over 2}{\partial u_1^2\over \partial x}=-
{\partial p_0\over \partial x}\,.
\label{equationsser4}
\end{equation}
\begin{equation}
{\partial \over \partial x}(H_1+u_1^2)=0\,.
\label{equationsser5}
\end{equation}
We keep the first order term ($\epsilon
u_1^2$) in the second equation of (\ref{equationsser2}) derived
in zero approximation over $\epsilon$. We do this because
it has been shown that in this case the analytical results on
the shock layer structure are more consistent with the results
of numerical simulations (see Hayes \& Probstein 1959;
Lunev 1975 and below).

Equations (\ref{equationsser1}) and (\ref{equationsser5})
are integrable by quadrature:
\begin{equation}
u_0=u_0(\psi)\,,\,\,\,\,\,\,\, p_0=p_0(x)\,,
\label{quadrsolution1}
\end{equation}
\begin{equation}
H_0=H_0(\psi)\,,\,\,\,\,\,\,\,
\rho_0={2\gamma\over (\gamma+1)}\,{p_0(x)\over H_0(\psi)}\,,
\label{quadrsolution2}
\end{equation}
\begin{equation}
u_1^2= -2\int_0^x
{dp_0(x)\over dx}{dx\over \rho_0(x,\psi)} +u^2_*(\psi)\,,
\label{quadrsolution3}
\end{equation}
\begin{equation}
y_0={1\over x}\int_0^{\psi}{d\psi\over \rho_0 (x,\psi)
[u_0^2(\psi)+\epsilon
u_1^2(x,\psi)]^{1/2}}+y_*(x)\,,
\label{quadrsolution4}
\end{equation}
\begin{equation}
v_0=u_0(\psi){\partial y_0(x,\psi)\over \partial x}\,,
\label{quadrsolution5}
\end{equation}
\begin{equation}
p_1=-\int_0^{\psi}{\partial v_0(x,\psi)\over \partial x}\,
{d\psi\over x} +p_*(x)\,,
\label{quadrsolution6}
\end{equation}
\begin{equation}
H_1=-u_1^2(x, \psi) +H_*(\psi)\,,
\label{quadrsolution7}
\end{equation}
\begin{equation}
\rho_1=\rho_0(x,\psi)\left[{p_1(x,\psi)\over p_0(x)}-
{H_1(x,\psi)\over H_0(\psi)}\right]
\label{quadrsolution8}
\end{equation}
where $u_0(\psi)$, $p_0(x)$, $H_0(\psi)$, $u_*(\psi)$, $y_*(x)$,
$p_*(x)$, and $H_*(\psi)$ are arbitrary functions that can be
determined from the boundary conditions at the shock wave and
at the contact surface.

Since $y=0$ and $\psi=0$ at the contact surface, from equation
(\ref{quadrsolution4}) we have $y_*(x)=0$.

\subsubsection{Zero approximation}

In the series (\ref{series1}) and (\ref{series2}) we keep only
the first terms marked by 0 (zero approximation).
To find the values $u_0$,
$p_0$, $H_0$, and $\psi_{0,s}$ in this approximation we can take that
the shock wave coincides with the contact plane, and
$\alpha$ is equal to $\pi/2$. From equations (\ref{behidshock1}),
(\ref{behidshock3}), and (\ref{behidshock4}) we have the following
boundary conditions:
\begin{equation}
u_{0,s}=\sin\varphi\,,\,\,\,\,p_{0,s}={\cos^2\varphi\over r_s^2}\,,
\,\,\,\,H_{0,s}=\cos^2\varphi\,.
\label{zerosolution1}
\end{equation}
From equation (\ref{psishock2}), we obtain
\begin{equation}
\psi_{0,s}=\int_0^x{xdx\over (1+x^2)^{3/2}}=1-{1\over (1+x^2)^{1/2}}\,.
\label{psist}
\end{equation}
Below, instead of $\psi$ we use the new coordinate $t$ which is
the length along the shock wave, measured from the binary axis to
the point where the stream line $\psi$ intersects the shock (see Fig.~2).

At the shock wave the value of $t$ in the zero approximation
is equal to $x$, and substituting $t$ for $x$ into equation
(\ref{psist}) we have the following connection between $\psi_0$
and $t$:
\begin{equation}
\psi_0=1-{1\over (1+t^2)^{1/2}}\,.
\label{psist2}
\end{equation}

In the new independent variables $t$ and $x$ ($0\leq t\leq x$),
equations (\ref{quadrsolution1}), (\ref{quadrsolution2}), and
(\ref{zerosolution1}) yield
\begin{equation}
u_0(t)=\sin \varphi (t)= {t\over (1+t^2)^{1/2}}\,,
\label{solutionfirst1}
\end{equation}
\begin{equation}
p_0(x)={\cos^2\varphi (t)|_{t=x}\over r_s^2(x)}=
{1\over (1+x^2)^2}\,,
\label{solutionfirst2}
\end{equation}
\begin{equation}
H_0(t)=\cos^2\varphi(t)={1\over 1+t^2}\,,
\label{solutionfirst3}
\end{equation}
\begin{equation}
\rho_0(x,t)={2\gamma\over (\gamma+1)}\,{p_0(x)\over H_0(t)}
={2\gamma\over (\gamma+1)}\,{1+t^2\over (1+x^2)^2}\,,
\label{solutionfirst4}
\end{equation}
where $r_s(x)=(1+x^2)^{1/2}$ [see equation (\ref{rs})].

From equations (\ref{quadrsolution4}) and (\ref{psist2})
we have the coordinate $y_0$ as a function of $x$ and $t$:
\begin{eqnarray}
y_0(x,t)={(\gamma+1)\over 2\gamma}\,{(1+x^2)^2\over x}
\int_0^t{dt\over (1+t^2)^2}
\nonumber\\
={(\gamma+1)\over 4\gamma}\,{(1+x^2)^2\over x}\, J(t)\,,
\label{solutionfirst5}
\end{eqnarray}
where
\begin{equation}
J(t)=\arctan t +{t\over 1+t^2}\,.
\label{Jt}
\end{equation}

From equations (\ref{quadrsolution5}), (\ref{solutionfirst1}),
and (\ref{solutionfirst5}), we obtain
\begin{equation}
v_0(x,t)={(\gamma+1)\over 4\gamma}\,{t\,J(t)\over (1+t^2)^{1/2}}\,
{(1+x^2)\,(3x^2-1)\over x^2}\,.
\label{v0}
\end{equation}

Since $t=x$ at the shock wave, from equations (\ref{series1})
and (\ref{solutionfirst5}) we obtain the equation of
the shock shape [see equation (\ref{shockshape})]
in the zero approximation (index $z$):
\begin{eqnarray}
y=y_s^z(x)=\epsilon y_0(x,t)|_{t=x}
={(\gamma-1)\over 4\gamma}\,{(1+x^2)^2\over x}J(x).
\label{solutionfirst6}
\end{eqnarray}

\subsubsection{Modified zero approximation}
In the modified zero approximation we take into account
only the $\epsilon$ correction ($\epsilon u_1^2$) to $u^2$.
In this approximation we also have $\alpha=\pi/2$, and from
equations (\ref{behidshock1}) and (\ref{behidshock4}) the boundary
conditions are $u_{1,s}=0$ and $H_{1,s}=0$ for $u_1$ and $H_1$,
respectively. Using these boundary conditions, from equations
(\ref{quadrsolution3}), (\ref{quadrsolution7}),
(\ref{solutionfirst2}), and (\ref{solutionfirst4})
we obtain
\begin{eqnarray}
u_1^2(x,t)=-2\int_t^x{dp_0\over dx}\,{dx\over \rho_0}\nonumber\\
=-{(\gamma+1)\over \gamma}\,{2\over (1+t^2)}\,
\ln {(1+t^2)\over (1+x^2)}\,,
\label{solutionfirstmod1}
\end{eqnarray}

\begin{equation}
H_1(x,t)=-u_1^2(x,t)\,.
\label{solutionfirstmod11}
\end{equation}

From equations (\ref{quadrsolution4}), (\ref{psist2}),
and (\ref{solutionfirstmod1}) the equation of
the shock shape in the modified zero approximation
(index $mz$) can be written as
\begin{eqnarray}
y=y_s^{mz}(x)=
{(\gamma-1)\over 2\gamma}\,{(1+x^2)^2\over x}
\int_0^x{t\,dt\over (1+t^2)^2}\nonumber\\
\times\left[t^2-{2(\gamma-1)\over \gamma}
\,\ln{(1+t^2)\over (1+x^2)}\right]^{-1/2}.
\label{solutionfirstmod2}
\end{eqnarray}

\subsubsection{First approximation}
We calculate now the shock layer parameters where
all $\epsilon$ corrections in the series
(\ref{series1}) and (\ref{series2}) are included
(first approximation).

Equations (\ref{behidshock1}), (\ref{behidshock3}),
(\ref{behidshock4}), and (\ref{shockshapealpha}) yield
the boundary conditions for $p_1$, $u_1$, and
$H_1$ near and behind the shock in the form:
\begin{equation}
p_{1,s}(x)=-{1\over (1+x^2)^2}\,,\,\,\,\,\,
u_{1,s}=0\,,\,\,\,\,\,H_{1,s}=0\,,
\label{boundcondfirstapp}
\end{equation}

Using the boundary conditions (\ref{boundcondfirstapp}),
from equations (\ref{quadrsolution6}), (\ref{quadrsolution7}),
and (\ref{v0}) we obtain
\begin{eqnarray}
p_1(x,t)=-{(\gamma+1)\over 2\gamma}\,
{(1+3x^4)\over x^4}
\int_t^x{t^2J(t)\,dt\over (1+t^2)^2}
-{1\over (1+x^2)^2}\nonumber\\
=-{(\gamma+1)\over 8\gamma}\,{(1+3x^4)\over x^4}[G(x)-G(t)]
-{1\over (1+x^2)^2}\,,
\label{solution1corr1}
\end{eqnarray}
\begin{equation}
H_1(x,t)=-u_1^2(x,t)\,,
\label{solution1corr2}
\end{equation}
where
\begin{equation}
G(x)=\arctan^2x-{2x\over (1+x^2)}\arctan x
-{2+3x^2\over (1+x^2)^2},
\label{solution1corr3}
\end{equation}
and $u_1(x,t)$ is given by equation (\ref{solutionfirstmod1}).

Equations (\ref{quadrsolution8}), (\ref{solutionfirst2}),
(\ref{solutionfirst3}), (\ref{solutionfirstmod1}),
(\ref{solution1corr1}), and (\ref{solution1corr2}) yield
\begin{eqnarray}
\rho_1(x,t)=-{4(1+t^2)\over (1+x^2)^2}
\left\{{(1+3x^4)(1+x^2)^2\over 16\,x^4}
[G(x)-G(t)]+
\ln{(1+t^2)\over (1+x^2)}\right\}.
\label{solution1corr4}
\end{eqnarray}

From the first equation of (\ref{equfinaldl1}) we have the equation
of the shock shape in the first approximation (index $f$):
\begin{equation}
y=y_s^f(x)={1\over x}\int_0^x{t\,dt\over (\rho_0+\epsilon \rho_1)\,
(u_0^2+\epsilon u_1^2)^{1/2}\,(1+t^2)^{3/2}}\,.
\label{solution1corr5}
\end{equation}

\subsection{The shock structure}
The shock shape in different approximations is given by
equations (\ref{solutionfirst6}), (\ref{solutionfirstmod2}), and
(\ref{solution1corr5}). From these equations it follows that the
dimensionless distance from the shock wave to the contact plane at
the binary axis ($x=0$) is
\begin{equation}
y_s^z(0)={\epsilon\over (1+\epsilon)}\,,
\label{shockstruc1}
\end{equation}
\begin{equation}
y_s^{mz}(0)={\epsilon\over (1-3\epsilon)}
\left[1-2\left({\epsilon\over 1+\epsilon}\right)^{1/2}\right]\,,
\label{shockstruc2}
\end{equation}
\begin{equation}
y_s^f(0) ={y_s^{mz}(0)\over (1-\epsilon)}
\label{shockstruc3}
\end{equation}
in the zero, modified zero, and first approximations,
respectively. The detachment of the shock wave and the contact
plane was calculated numerically by Lebedev \& Savinov (1969) and
Savinov (1975), and for $1.2<\gamma<1.8$, i.e., for
$0.09<\epsilon<0.29$, its dependence on $\epsilon$ at $x=0$ was
fitted by
\begin{equation}
y_s^n(0)=(0.365\,\epsilon+0.035)\,. \label{shockstruc4}
\end{equation}
Here, we briefly discuss the shock shapes calculated in different
approximations for the case of $\gamma=5/3$, which may have a
special interest for colliding winds in massive binaries (see
below).

For $\gamma={5\over 3}$ and $\epsilon={1\over 4}$ from equations
(\ref{shockstruc1})-(\ref{shockstruc4}) we have $y_s^z(0)=0.2$,
$y_s^{mz}(0)\simeq 0.106$, $y_s^f(0)\simeq 0.141$, and
$y_s^n(0)\simeq 0.126$. We can see that the $\epsilon$ corrections
in the series (\ref{series1}) and (\ref{series2}) make more
precise the calculations of $y_s(0)$.

The shock shapes in different approximations and the results of
numerical simulations (Lebedev \& Savinov 1969; Savinov 1975) are
plotted in Figure~3 in the dimensionless coordinates $x$ and $y$
for $0\leq x\leq 1$. The shock shape in the modified zero
approximation is the most consistent with the results of numerical
simulations fulfilled by Lebedev \& Savinov (1969) and Savinov
(1975).  Figure~3 shows that at large distances ($x\gtrsim 0.9$)
from the axis the shock shape in the first approximation where all
$\epsilon$ corrections are included into consideration differs
from the numerical result even more than the shock shape in the
simple zero approximation where all $\epsilon$ corrections are
ignored, and this difference sharply increases with increase of
$x$. Such a paradoxical behavior of the shock shape calculated in
the first $\epsilon$ approximation for $\gamma -1\sim 1$ is known
and has been mentioned in many papers (for a review, see Lunev
1975). It is connected with the following. The $\epsilon$
correction of the gas pressure (\ref{solution1corr1}) is negative
and rather slowly decreases with increases of $x$. At $x\simeq 1$
the absolute value of this correction is comparable with the gas
pressure in the zero approximation (\ref{solutionfirst2}), and the
used analytical method where all $\epsilon$ corrections have to be
small is not applicable. In fact, the paradoxical behavior of our
solution at $x\simeq 1$ is because for $\gamma=5/3$ the value of
$\epsilon=1/4$ is not really small enough. Therefore, to escape
the difficulty at $x\simeq 1$ and to improve the solution in the
zero approximation it was suggested the modified zero
approximation where only a part of $\epsilon$ corrections are
included (Freeman 1956, 1958). The accuracy of the last
approximation for $\gamma=5/3$ is $\sim 10-20$\% that is higher
than the accuracy of the zero approximation (see Figure~3). Since
in the method developed by Hayes \& Probstein (1959) and Chernyi
(1961) and used in our calculations the value of $\epsilon$ is
considered as a small parameter the accuracy of our results has to
increases as $\gamma$ approaches unity. We hope to make a detail
comparison between the analytical and numerical results for
different values of $\gamma$ elsewhere.

\section{Discussion}
We have considered in this paper the collision of two identical
hypersonic steady winds in a binary system. We neglect the
rotation of the binary and its components. This is expedient
because the accuracy of our calculations is not higher than $\sim
10$\% while in most observed massive binaries, for which we are
going to use our results in future, typical stellar wind
velocities exceed typical orbital velocities by a factor of $\sim
10$; thus, the orbital rotation doesn't distort substantially
(more than $\sim 10$\%) the gas flows out to distances equal to
several orbital separations. We assume that the winds from the
binary components are radial and spherically symmetrical. In this
case the contact surface that separates the gas emanated from the
different stars is known before calculations from a reflection
symmetry about the midplane of the stars and coincides with the
midplane (see Fig.~1). The two shock layers between the contact
surface and the shock waves are the same because of the reflection
symmetry, and therefore, it is necessary to consider only the
properties of one of them. The problem is solved analytically,
using the method in which the ratio of the gas density ahead of
the shock to that behind it is considered as a small parameter,
$\epsilon \ll 1$ (e.g., Bairamov et al. 1990; Usov 1992 and
references therein). We have found the shape of the shock wave and
the gas parameters in the shock layer. We have done this for
$\gamma=5/3$ ($\epsilon=1/4$) in the following three
approximations. In the zero approximation we keep only the first
terms in the $\epsilon$-series (\ref{series1}) and (\ref{series2})
for the shock layer parameters. In the modified zero approximation
we keep additionally only the second term ($\epsilon u^2_1$) in
the $\epsilon$-series of $u^2$ (\ref{series1}). In the first
approximation all $\epsilon$ corrections in the series
(\ref{series1}) and (\ref{series2}) are included. We have shown
that the shock shape calculated in the modified zero approximation
is the most consistent with the results of numerical simulations
(see Figure~3). The expected accuracy of the shock layer
parameters calculated in this approximation is $\sim 10-20$\%.
These parameters may be used for calculations of emission from the
shock layers, including the X-ray line profiles.

The value of $\gamma=5/3$ has a special interest for our problem
because it may relate to the hot gas in the shock layers in
massive binaries if these binaries are not too wide. Indeed, for
massive binaries the gas temperature in the shock layers is very
high ($\sim 10^6-10^7$~K or even higher) if the distance from the
axis is not too large ($x\lesssim 1$). In the case of local
thermodynamic equilibrium at such temperatures light elements (H,
He, C, etc.) are practically totally ionized while atomic nuclei
of more heavy elements (Ne, Mg, Si, S, and Fe) have a few bound
electrons, and these strongly ionized, heavy ions are responsible
for the X-ray emission lines observed from massive binaries (e.g.,
Henley et al. 2005). It is know that in WR winds hydrogen is
essentially absent, and helium predominates. The relative
abundances of neon and more heavier elements is $\lesssim
3.5\times 10^{-3}$ (Dessart et al. 2000). Taking into account that
these heavy elements are highly ionized in the shock layer, we can
see that the deviation of the gas pressure from the pressure of
totally ionized plasma is not larger than $\sim 1$\%. In this
case, a totally ionized plasma is a good approximation for the
state of the hot gas, and $\gamma$ is nearly 5/3. The same
approximation may be applicable even better for the hot gas of the
shock layers in OB binaries where the wind composition doesn't
differ significantly from the solar abundances, and hydrogen
predominates (Anders \& Grevesse 1989). Therefore, the equation of
state with $\gamma=5/3$ is usually used for both numerical and
analytical studies of colliding winds in massive binaries (e.g.,
Luo et al. 1990; Stevens et al. 1992; Usov 1992). The question is
there is or not local thermodynamic equilibrium in the shock
layers. Recently, using an archived {\it Chandra} HETGS X-ray
spectrum of the WR+O colliding wind binary $\gamma^2$ Velorum it
was found evidence that the Mg XI emission originates from hotter
gas closer to the O star than the Si XIII emission, which suggests
that ionization of these elements may be non-equilibrium (Henley
et al. 2005). But, this cannot significantly change the equation
of state for the hot gas in the shock layer because the abundances
of Mg and Si are very small. In another X-ray observations by {\it
Chandra} presented by Pollock et al. (2005) for the wide
Wolf-Rayet binary WR 140 there is evidence that the temperatures
of ions and electrons in the shock layers are different. Such a
difference of the ion and electron temperatures is expected for
wide binaries. The point is that the postshock temperature for
ions is much higher that the same for electrons (Zel'dovich \&
Raizer 2002). In the process of the gas outflow the electrons are
heated by collisions with the ions, and their temperatures are
equalizing if the gas density in the shock layer is high enough,
i.e., if the binary is not too wide (Usov 1992). For two identical
colliding winds the condition of the temperature equalization may
be written as
\begin{equation}
2D < 10^{14}\left({\dot M\over 10^{-6}\,M_\odot\,{\rm
yr}^{-1}}\right)\left({V^\infty\over 10^3\, {\rm
km~s}^{-1}}\right)^{-5}\,\,{\rm cm}\,. \label{equalization}
\end{equation}
We can see that the restriction (\ref{equalization}) on the binary
separation $2D$ is rather hard if the wind velocity is large. For
WR 140 the terminal velocities of the WR and O winds are high
($\sim 3000~{\rm km~s}^{-1}$), and therefore, the electron and ion
temperatures are not equalized in the shock layers. If for a
binary the condition (\ref{equalization}) is satisfied the
one-temperature equation of state (\ref{ptotal}) is correct for
the hot gas, and $\gamma$ is nearly 5/3.

It is known that the gas flow in the region of the wind-wind
interaction region may be unstable, especially if the cooling
timescale is not small in comparison with the flow timescale
(e.g., Usov 1991; Stevens et al. 1992; Walder \& Folini 2000). We
are planning to take into account radiative cooling of the hot
gas, and then, to study stability of our new solution for the
shock layer.

\begin{acknowledgements}
We thank the anonymous referees for many helpful suggestions that
improved the paper considerably. The research was supported by the
Israel Science Foundation of the Israel Academy of Sciences and
Humanities.
\end{acknowledgements}

\clearpage

\begin{figure}
\plotone{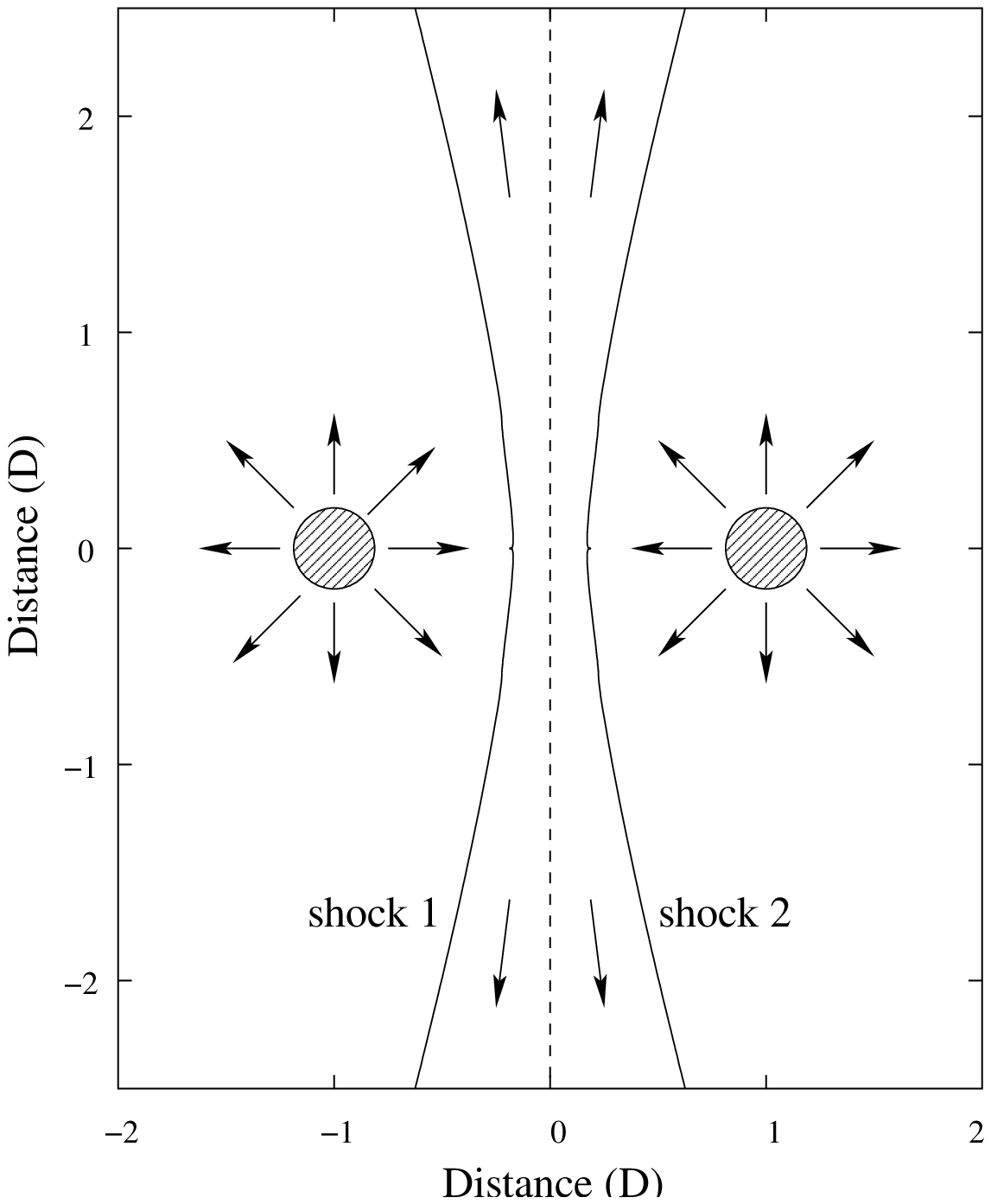} \caption{Sketch of a colliding-wind binary system
in which the components (hatched) are the sources of identical
hypersonic winds. The hot gas is bounded by two shocks (solid lines).
The gas emanated from the different stars is separated by a contact
surface (dashed line). The directions of gas flow are shown by arrows.}
\label{SketchBinary}
\end{figure}
\begin{figure}
\plotone{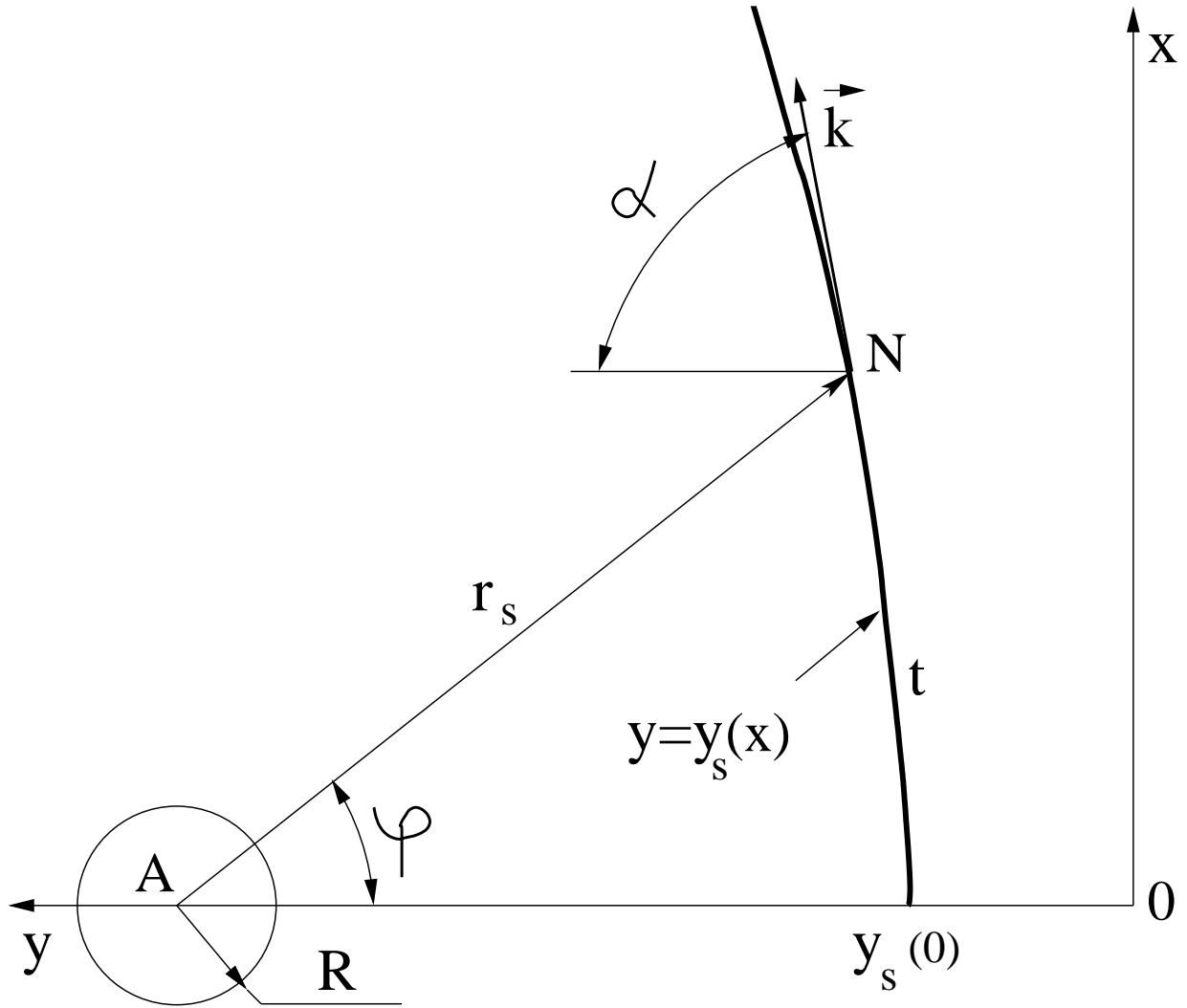} \caption{The shock wave (thick solid line)
with the equation $y=y_s(x)$ in the coordinates $x$ and $y$ where
the axis $x$ belong to the contact plane ($y=0$), and the axis $y$
coincides with the binary axis ($x=0$). A-point is the center of the star,
and N is the point where the radial current line AN intersects the shock
wave. The vector $\vec k$ is the tangent to the shock wave at the point N,
and $t$ is the coordinate of the point N which is the length along the
shock wave, measured from the binary axis to this point.}
\label{Sketchangles}
\end{figure}
\begin{figure}
\plotone{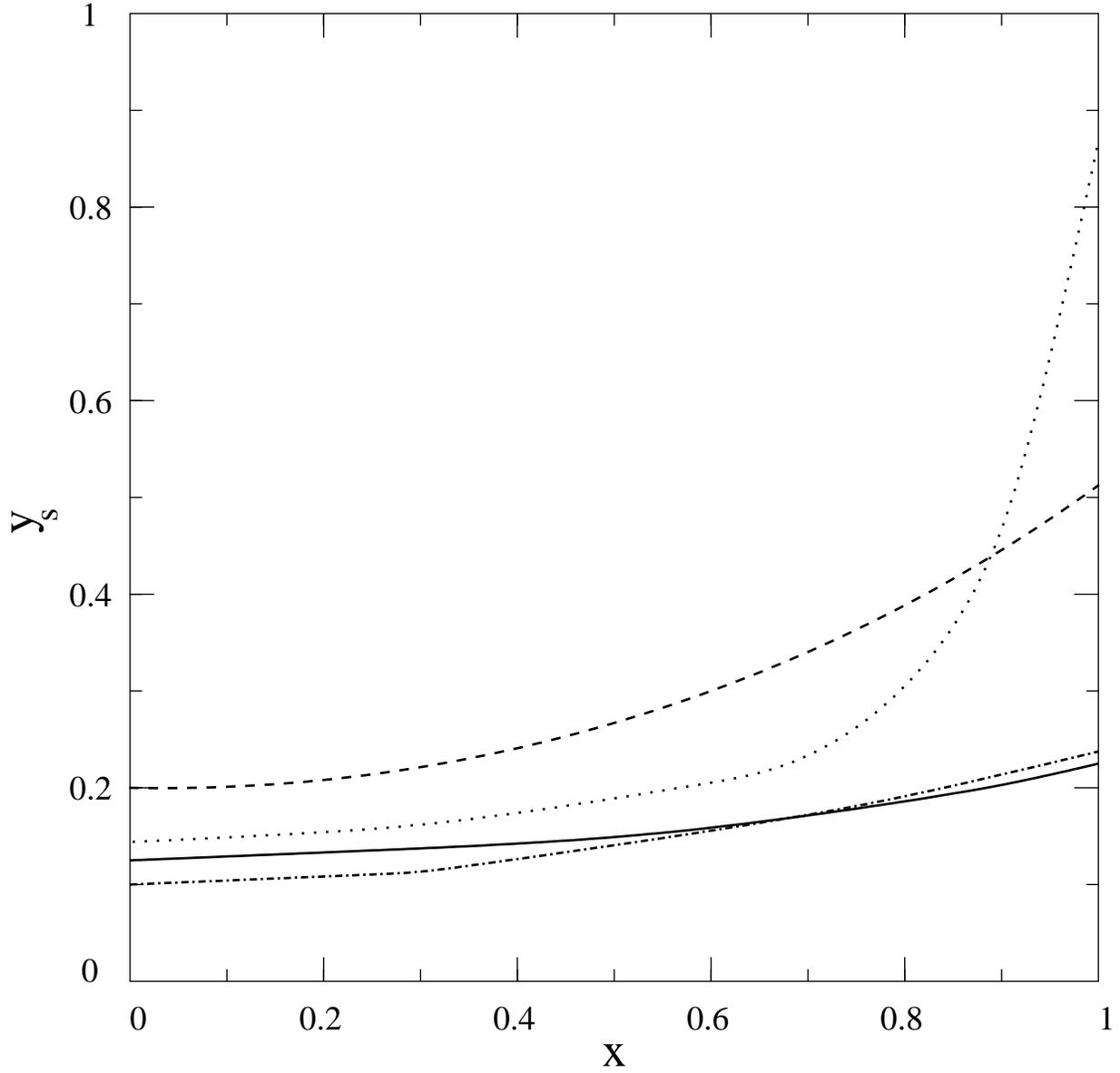} \caption{The shock shape in the zero (dashed curve),
modified zero (dot-dashed curve), and first (dotted curve) approximations
in the dimensionless coordinates $x$ and $y$ where the axis $x$ belong
to the contact plane, and the axis $y$ coincides with the binary
axis. The result of numerical simulations is represented by the solid
curve.}
\label{Sketchshocks}
\end{figure}

\end{document}